\documentstyle[11pt,fleqn]{article}

\begin{document}


\hfill{\sl preprint - UTF 401 \\ hep-th/9705077 }
\par
\bigskip
\par
\rm


\par
\bigskip
\begin{center}
\bf
\LARGE
$\zeta$-function regularization and one-loop renormalization of 
field fluctuations in curved space-times
\end{center}
\par
\bigskip
\par
\rm
\normalsize



\begin{center}\Large

Valter Moretti \footnote{e-mail:
\sl moretti@science.unitn.it} 
and Devis Iellici\footnote{e-mail:
\sl iellici@science.unitn.it} 

\end{center}



\begin{center}
\large
\smallskip

 Dipartimento di Fisica, Universit\`a di Trento
\\and

\smallskip

 Istituto Nazionale di Fisica Nucleare,\\
Gruppo Collegato di Trento,\\ 38050 Povo (TN)
Italia

\end{center}\rm\normalsize



\par
\bigskip
\par
\hfill{\sl May 1997}
\par
\medskip
\par\rm



\begin{description}
\item{Abstract:\\ 
\it  A method to regularize and renormalize the fluctuations of a
quantum field in a curved background in the $\zeta$-function approach
is presented. The method produces finite quantities directly  and
finite scale-parametrized  counterterms at most. These finite
couterterms are related to the presence of a particular pole of the
effective-action $\zeta$ function as well as to the heat kernel
coefficients. The method is checked in several examples obtaining
known or reasonable results. Finally, comments are given for as it
concerns the recent proposal by Frolov et al. to get the finite
Bekenstein-Hawking entropy from  Sakharov's induced gravity theory.}
\par
\end{description}
\rm


\noindent{\sl PACS number(s):\hspace{0.3cm}
04.62.+v, 04.70.Dy}
\par
\bigskip
\rm



\section*{Introduction}

This  letter is devoted to develop a possible $\zeta$-function approach to
regularize and renormalize the averaged square field
$\langle\phi\sp{2}(x)\rangle$ in a general curved background. That
quantity has been studied by several authors \cite{bd} because its
knowledge is an  important step to proceed with the point-splitting
renormalization of the stress tensor and also  due to its importance
in cosmological theories. The knowledge of the  value
of $\langle\phi\sp{2}(x)\rangle$ is also necessary to get the
renormalized Hamiltonian from the renormalized stress tensor in non
minimally coupled theories and this could be important dealing with
thermodynamical considerations
\cite{frofurzel97a,frofurzel97b,frofurzel97c,Zmoretti}. Recently, it
has also been studied in relation to the  black hole entropy physics
\cite{frofurzel97a,frofurzel97b,frofurzel97c}.

Let us consider a generic Euclidean field theory in curved background
${\cal M}$ corresponding to the  Euclidean action 
\begin{eqnarray}
S\sb{A}[\phi] = -\frac{1}{2}\int\sb{\cal M} d\sp{4}x \sqrt{g}\: 
\phi A \phi, 
\end{eqnarray}
where Euclidean motion operator $A$ is supposed self-adjoint
and positive-definite. The  local $\zeta$ function related to the
operator $A$, $\zeta(s,x|A)$, can be defined as the 
analytic continuation of the series \cite{hawking,zerbinil}
\begin{eqnarray}
\zeta(s,x|A) = {\sum\sb{n}}'
 \lambda\sb{n}\sp{-s} \phi\sb{n}\sp{*}(x)\phi\sb{n}(x) 
\label{zeta}.
\end{eqnarray}
where $\{\phi\sb{n}(x)\}$ is a complete set of normalized eigenvectors
of $A$ with eigenvalues $\lambda\sb{n}$, and in the sum the null
eigenvectors are omitted. In Eq.(\ref{zeta}), the right hand side is
supposed to be analytically continued in the whole $s$-complex plane.
Indeed, it is well-known that the $\zeta$ function is a meromorphic
function with, at most, simple poles situated at $s=1$ and $s=2$ in
case of a four dimensional spacetime. We are working formally with a
discrete spectrum, but all considerations can be trivially extended to
operators with continuous spectrum. The most importance of the $\zeta$
function is that the derivative of such a function evaluated at $s=0$
defines a regularization of the one-loop effective action:
\begin{eqnarray*}
S\sb{\mbox{\scriptsize eff}}[\phi,g]=\frac{1}{2}\frac{d}{ds}
\zeta(s,x|A/\mu\sp{2})
\end{eqnarray*}
where $\mu$ is an arbitrary parameter with the dimensions of a  mass
necessary from dimensional considerations. Recently, the method has
been extended in order to define a similar $\zeta$-function
regularization directly for the renormalized stress tensor
\cite{Zmoretti}.

The usual approach to compute the field fluctuation by 
the $\zeta$-function techniques, leads to
the naive definition for the one-loop averaged square field
\begin{eqnarray}
\langle\phi\sp{2}(x)\rangle := \zeta(s,x|A/\mu\sp{2})|\sb{s=1} 
\label{naive},
\end{eqnarray}
This definition follows directly from (\ref{zeta}) taking account of
the spectral decomposition of the two-point function. Anyhow, barring
exceptional situations (e.g. see \cite{moretticonic}), this definition
is not available because the presence of a pole at $s=1$ in the
analytically continued $\zeta$ function and further infinite
subtraction procedures seem to be necessary. Conversely, in the cases
of the effective action and  stress tensor regularization, the
$\zeta$-function approach leads naturally to the  complete
cancellation of divergences maintaining the finite 
$\mu$-parameterized counterterms physically necessary \cite{Zmoretti}.
We shall see shortly that, also in the case of
$\langle\phi\sp{2}(x)\rangle$, it is possible to improve the
$\zeta$-function approach to get the same features: cancellation of
all divergences  maintenance of the finite $\mu$-parameterized
counterterms.

Finally we shall check the approach in some cases of physical
interest, producing some comments on the recent proposal to explain
the Bekenstein-Hawking entropy in the framework of Sakharov's induced
gravity \cite{frofurzel97a,frofurzel97b}.

\section{The general approach}

Let us define the $\zeta$ function of $\langle\phi\sp{2}(x)\rangle$. The
way we will follow is very similar to that followed in \cite{Zmoretti}
to compute the $\zeta$ function of the stress tensor.
The one-loop effective action is given as
\begin{eqnarray}
S\sb{\scriptsize \mbox{eff}}[g\sb{ab}] = 
\ln\int{\cal D}\phi\:e\sp{S\sb{A}}=
\ln \int {\cal D}\phi\: e\sp{S\sb{A}|\sb{m=0} - 
\frac{1}{2} \int d\sp{4}x \sqrt{g}\:m
\sp{2}
\phi\sp{2}}
\label{effective}.
\end{eqnarray}
Let us temporarily suppose to go ``off-shell'' as far as the field
mass is concerned, namely let us put 
$m\sp{2} \rightarrow m\sp{2}(x)$,
where $m\sp{2}(x)$ is a general smooth function. Then we have
\begin{eqnarray}
\langle\phi\sp{2}(x)\rangle = -\frac{2}{\sqrt{g(x)}}\:
\frac{\delta S\sb{\scriptsize \mbox{eff}}}{\delta 
m\sp{2}(x)}|\sb{m\sp{2}(x) = m\sp{2}}, \label{formal}
\end{eqnarray}
where $m\sp{2}$ is the actual value of the field mass.
{From} the $\zeta$-function regularization, we can formally write
\begin{eqnarray}
\langle\phi\sp{2}(x)\rangle \mbox{``}=\mbox{''} \:
\frac{1}{\sqrt{g(x)}}
\frac{\delta \:\:\:\:}{\delta
m\sp{2}(x)}|\sb{m\sp{2}(x) = m\sp{2} }
\frac{d\:\:}{ds}|\sb{s=0}
\zeta(s|A/\mu\sp{2}) 
\nonumber
\end{eqnarray}
or, supposing to be possible to interchange the order of the derivatives
\begin{eqnarray}
\langle\phi\sp{2}(x)\rangle \mbox{``}=\mbox{''} \: 
\frac{d\:\:}{ds}|\sb{s=0}
\frac{1}{\sqrt{g(x)}}
\frac{\delta \zeta(s|A/\mu\sp{2})}{\delta
m\sp{2}(x)}|\sb{m\sp{2}(x) = m\sp{2}}.\nonumber
\end{eqnarray}

A definition of the functional derivative of the $\zeta$ function is
now necessary. We get such a definition following the same way as in
\cite{Zmoretti} for the case of stress tensor. Let us first notice
that one can quite simply prove (see Appendix in \cite{Zmoretti} where
proofs are carried out for similar identities)
\begin{eqnarray}
\frac{\delta \lambda\sb{n}}{\delta m\sp{2}(x)} =
-\phi\sb{n}\sp{*}(x)\phi\sb{n}(x)\sqrt{g(x)}.
\label{dl}
\end{eqnarray}
Then, we can {\em define}
\begin{eqnarray}
\frac{\delta \zeta(s|A/\mu\sp{2})}{\delta
m\sp{2}(x)} := {\sum\sb{n}}' 
\frac{\delta \lambda\sp{-s}\sb{n}}{\delta m\sp{2}(x)}
\end{eqnarray} 
where we suppose to analytically continue the right hand side as far
as is possible in the complex $s$ plane. As usual, the summation does
not include null eigenvalues. Then, employing (\ref{dl}) we have (notice
that the eigenvalues of $A/\mu\sp{2}$ are $\lambda\sb{n}/\mu\sp{2}$)
\begin{eqnarray}
\frac{\delta \zeta(s|A/\mu\sp{2})}{\delta
m\sp{2}(x)} =\sqrt{g(x)}\: \frac{s}{\mu\sp{2}}\: 
\zeta(s+1,x | A/\mu\sp{2}). \nonumber
\end{eqnarray}
Substituting this result in the right hand side of (\ref{formal})
we have our definition of the $\zeta$-renormalized 
$\langle\phi\sp{2}(x)\rangle$
\begin{eqnarray}
\langle\phi\sp{2}(x)\rangle := \frac{d\:\:}{ds}|\sb{s=0}\: 
\frac{s}{\mu\sp{2}}\:
\zeta(s+1,x|A/\mu\sp{2}) \label{phisquare},
\end{eqnarray}
where the $\zeta$ function is that evaluated at the actual value of the 
mass $m$.
The identity above can be written down also as
\begin{eqnarray}
\langle\phi\sp{2}(x)\rangle := \frac{d\:\:}{ds}|\sb{s=0}
Z(s|A/\mu\sp{2})\:  \nonumber,
\end{eqnarray}
where
\begin{eqnarray}
Z(s|A/\mu\sp{2})
 := \frac{s}{\mu\sp{2}}\:
\zeta(s+1,x|A/\mu\sp{2}) \nonumber
\end{eqnarray}
is the $\zeta$ {\em function of the field fluctuation}.

Notice that the simple pole at $s=0$ in $\zeta(s+1,x|A/\mu\sp{2})$,
whenever it exists, is canceled out by the factor $s$; moreover, when
$\zeta(s+1,x|A/\mu\sp{2})$ is regular at $s=0$, the  definition 
in (\ref{phisquare})
coincides with the naive one, Eq. (\ref{naive}).

It is worth while stressing that, in general, the scale $\mu$ can
remain into the final result. It represents the usual ambiguity due to
the remaining finite renormalization already found as far as the
effective action and the stress tensor are concerned \cite{Zmoretti}.
In general, the scale could remain into the final result whenever
another {\em fixed} scale is already present into the theory, e.g. the
mass of the field or the curvature of the spacetime. Actually, it can
disappear also in those cases provided particular conditions hold true
(see below).  The disappearance of the scale $\mu$ from the final
result is equivalent to the possibility of using  the definition
(\ref{naive}). Indeed, (\ref{phisquare}) can be also written as
\begin{eqnarray}
\langle\phi\sp{2}(x)\rangle := \frac{d\:\:}{ds}|\sb{s=0}\: s\:
\zeta(s+1,x|A) + \
 s \zeta(s+1,x|A)|\sb{s=0} \ln \mu\sp{2}.
\label{phisquare2}
\end{eqnarray}
We see that $\ln \mu\sp{2}$ disappears if and only if $\zeta(s+1,x|A)$
is analytic at $s=0$, namely, if and only if we can use the definition
(\ref{naive}).

Finally, it is  intersting to  investigate the relation between the
presence of a pole in the $\zeta$ function and the heat-kernel
coefficients. From the well-known asymptotic expansion of the local
heat kernel on manifolds without boundary
\begin{eqnarray}
K\sb{t}(x|A/\mu\sp{2})\simeq \frac{1}{(4\pi)\sp{D/2}}\sum\sb{j=0}
\sp{\infty}k\sb{j}(x|A/\mu\sp{2}) t\sp{j-D/2},
\nonumber
\end{eqnarray}
it is easy to show that the $\zeta$ function has a pole in $s=1$
if and only if the dimension $D$ is even and the coefficient
$k\sb{\frac{D}{2}-1}(x|A)$ is non-vanishing.
Therefore, in odd dimensions we can use the definition (\ref{naive})
of the fluctuations, while in even dimensions we can rewrite
the definition (\ref{phisquare}) as
\begin{eqnarray}
\langle\phi\sp{2}(x)\rangle :=\lim\sb{s\rightarrow 1}
\left[\zeta(s,x|A/\mu\sp{2})-\frac{k\sb{\frac{D}{2}-1}
(x|A/\mu\sp{2})}
{(4\pi)\sp{D/2}\Gamma(s)(s-1)}\right]+
\frac{\gamma}{(4\pi)\sp{D/2}}k\sb{\frac{D}{2}-1}(x|A/\mu\sp{2}),
\nonumber
\end{eqnarray}
where $\gamma$ is Euler's constant. In four dimensions we have that
$k\sb{1}=\frac{1}{6}(1-6\xi)R(x)-m\sp{2}$, and so it is clear that the
naive definition, Eq. (\ref{naive}), is available only in the massless
case with vanishing scalar curvature or in the massless conformal
coupling case, $\xi=1/6$.

\section{Simple applications and comments}

As a first application we consider a scalar field in Minkowski
space time. It is well known \cite{AAA95} that for a massless
field the corresponding $\zeta$ function can be considered
as vanishing, and so we pass directly to the massive case.
The local $\zeta$ function in four dimensions reads
\begin{eqnarray*}
\zeta(s,x|A/\mu\sp{2})=\frac{m\sp4(\mu/m)\sp{2s}}
{16\pi\sp{2}(s-1)(s-2)},
\end{eqnarray*}
which has a simple pole at $s=1$. Using Eq. (\ref{phisquare}) we obtain
\begin{eqnarray}
\langle \phi\sp{2}(x)\rangle=\frac{m\sp{2}}
{16\pi\sp{2}}\left[2\ln\frac{m}{\mu}-1\right].
\label{mink}
\end{eqnarray}

Then we  consider a scalar massless field in Minkowski spacetime
contained in a large box at the temperature $\beta$. The local $\zeta$
function is simply obtained (see \cite{hawking}) and reads
\begin{eqnarray}
\zeta(s,x|A/\mu\sp{2} ) = \frac{\sqrt{\pi}\mu\sp4}{(2\pi)\sp3}
\left( \frac{2\pi}{\beta \mu}\right)\sp{4-2s}
\frac{\Gamma(s-3/2)}{\Gamma(s)}\zeta\sb{R}(2s-3)\nonumber,
\end{eqnarray}
where $\zeta\sb{R}(s)$ is the usual Riemann zeta function. Notice that
no pole appears at $s=1$, hence we could also use the naive definition
(\ref{naive}) instead of (\ref{phisquare}). In both cases the result
is
\begin{eqnarray}
\langle\phi\sp{2}(x)\rangle\sb{\beta} = \frac{1}{12 \beta\sp{2}}.
\nonumber
\end{eqnarray}
This result is the same which follows from other approaches (e.g.,
subtracting the Minkowski massless zero temperature two-point 
function from the thermal one and performing the limit of coincidence
of the arguments).

Now we consider the Casimir effect due to two infinite parallel planes
on which the field is constrained to vanish, namely we consider a
massless scalar field in the Euclidean manifold $[0,L]\times R\sp3$.
In this case the local $\zeta$ function can be computed taking the
Mellin transform of the corresponding heat kernel, which is given,
e.g., in \cite{AAA95}. A straightforward computation yields
($0<x<L/2$)
\begin{eqnarray}
\zeta(s,x|A/\mu\sp{2})=\frac{L\sp{2s-4}\Gamma(2-s)}{16\pi\sp{2} 
\Gamma(s)}\left[2\zeta\sb{R}(4-2s)+\left(\frac{x}{L}\right)
\sp{2s-4}-2\zeta\sb{R}(4-2s,x/L)\right].
\end{eqnarray}
We see that the $\zeta$ function is regular at $s=1$ and so the
fluctuations can be computed in the naive way: 
\begin{eqnarray}
\langle\phi\sp{2}(x)\rangle\sb{\beta}
&=&\frac{1}{48L\sp{2}}-\frac{1}{8\pi\sp{2}L\sp{2}}
\left[\zeta\sb{R}(2,x/L)-\frac{L\sp{2}}{2x\sp{2}}\right]\nonumber\\
&=&\frac{1}{48L\sp{2}}\left[1-3\mbox{csc}\sp{2}
\frac{\pi x}{L}\right]\nonumber\\
&\sim&-\frac{1}{16\pi\sp{2}x\sp{2}}+{\cal O}(x\sp0),
\nonumber
\end{eqnarray}
in agreement with the known result, see, e.g., \cite{fullingbook}.

As the simplest application on a curved background we may consider the
field fluctuation of a massless conformally coupled scalar field on
the open static Einsten universe with spatial section metric
$a\sp{2} [dX\sp{2} + \sinh\sp{2} X (d\theta\sp{2} +
 \sin\sp{2}\theta d\varphi\sp{2})] $.
The modes can be built up following \cite{bunch}.
A few calculations lead  to the thermal local $\zeta$
function
\begin{eqnarray}
\zeta(s,x|A) =
\frac{\sqrt{\pi}\mu\sp4}{2(2\pi)\sp{2}}
\left( \frac{2\pi}{\beta \mu}\right)\sp{4-2s}
\frac{\Gamma(s-3/2)}{\Gamma(s)}\zeta\sb{R}(2s-3)\nonumber. 
\end{eqnarray}
In practice, this is  the  $\zeta$ function in the Minkowski large box.
Hence we have similarly
\begin{eqnarray}
\langle\phi\sp{2}(x)\rangle\sb{\beta} = \frac{\pi}{12 \beta\sp{2}}.
\nonumber
\end{eqnarray}
and thus at zero temperature
$\beta\rightarrow +\infty$  and
$\langle\phi\sp{2}(x)\rangle\sb{\beta} \rightarrow 
\langle\phi\sp{2}(x)\rangle\sb{\scriptsize\mbox{vacuum}} = 0$ 
as well-known
\cite{bunch}.

Another interesting application is the case of a massive scalar field
near an idealized GUT cosmic string with deficit angle
$2\pi(1-1/\nu)$, whose local $\zeta$ function has been recently
computed \cite{massive} in the limit $mr\ll 1$,  $r$ being is the
distance from the core of the string. Keeping terms up to ${\cal
O}\left((mr)\sp{2}\right)$, we get the following expression for the
fluctuations
\begin{eqnarray}
\langle\phi\sp{2}(x)\rangle=\frac{1}{48\pi r\sp{2}}
\left\{\nu\sp{2}-1+6(mr)\sp{2}\left[(\nu-1)
\left(\gamma-\ln\frac{r\mu}{2}\right)+\ln\nu+
\frac{\nu}{2}\left(2\ln\frac{m}{\mu}-1\right)\right]\right\}.
\label{cosmic}
\end{eqnarray}
The above expression is different from the result obtained subtracting
the Minkowski value \cite{moreira95,massive}, and in fact
reduces to the Minkowski value, Eq. (\ref{mink}), rather than vanishing
when the conical singularity is removed, namely $\nu\rightarrow 1$.
We note that in the massless case both procedures give the same result,
since the massless $\zeta$ function is regular at $s=1$.

The expression of the fluctuations given in Eq. (\ref{cosmic}) may
also be interpreted as giving the fluctuations of a massive scalar
field in the Rindler space at temperature $T=1/\beta=\nu/2\pi$. This
is a consequence of the well-known correspondence between the  cosmic
string background and the the Rindler space, identifing the cone angle
$\beta=2\pi/\nu$ with the inverse temperature. In this case the expression
(\ref{cosmic}) is valid near the event horizon, $r=0$. 

It is interesting to use this expression to evaluate the
Bekenstein-Hawking entropy of a black hole in the framework of
Sakharov's induced gravity \cite{jacobson94}, as done by Frolov {\em
et al.} \cite{frofurzel97a,frofurzel97b}. In this appealing approach,
Einstein action arises as the low-energy limit of the effective action
of some quantum fields of large mass, $N\sb{s}$ {\em non minimally}
coupled scalar fields  and $N\sb{d}$ fermion fields. The
Bekenstein-Hawking entropy is identified with the entropy of these
fields propagating in the black hole background, which can be computed
using the conical manifold method corrected with a surface term:
\begin{eqnarray}
S\sp{BH}(\beta\sb{H})=\beta\sp{2}\partial\sb{\beta} 
F\sb{\beta}|\sb{\beta=\beta\sb{H}}
-\sum\sb{s} 2\pi\xi\sb{s} \int\sb{\Sigma}
 d\sigma\langle\phi\sp{2}\sb{s}\rangle,
\label{induced}
\end{eqnarray}
where $\beta\sb{H}$ is the Hawking temperature for which the conical
singularity is absent, $F\sb{\beta}$ is the free energy of the fields
propagating in the conical manifold and $\Sigma$ is the horizon.

We want to discuss that problem  very briefly within our approach
(related massless cases without to consider Sakharov's theory have
been recently discussed in \cite{moretticonic,hotta}). Approximating
the black-hole metric near the horizon with the Rindler metric, we can
compute the free energy of a single scalar field form the local $\zeta$
function as $F\sb{\beta}=-\beta\sp{-1}Z\sb{\beta}$,
\begin{eqnarray*}
Z\sb{\beta}=\int {\cal L}(x)\sqrt{g}d\sp4x=
\int \frac{1}{2}\frac{d}{ds}
\zeta(s,x|A/\mu\sp{2})|\sb{s=0}\sqrt{g}d\sp4x,
\end{eqnarray*}
where the $\zeta$ function is the one given in \cite{massive}.
Employing the result (\ref{cosmic}) at the Hawking temperature,
namely $\nu=1$, we get 
\begin{eqnarray}
S\sp{BH}(\beta\sb{H})=\frac{{\cal A}\sp{H}}{120\pi\epsilon\sp{2}}
\left[2 - 5 m\sb{s}\sp{2}\epsilon\sp{2}\ln(R/\epsilon)+
15\pi m\sb{s}\sp{2}\xi\sb{s}
\epsilon\sp{2}(2\ln(m\sb{s}/\mu)-1)\right],
\label{entropy}
\end{eqnarray}
where ${\cal A}\sp{H}$ is the area of the horizon, $\epsilon$ is a
minimal distance from the horizon (`horizon cut off'  \cite{thooft})
and $R$ is a large radius needed to control the volume divergence. As
before, we have considered only terms up to order ${\cal
O}\left((m\sb{s}r)\sp{2}\right)$, which is justified near the horizon.

It is clear from Eq. (\ref{entropy}) that the surface terms
$\int\sb{\Sigma} d\sigma\langle\phi\sp{2}\sb{s}\rangle$ in Eq.
(\ref{induced}) cannot cancel out the {\em horizon} divergences (as
$\epsilon\rightarrow 0$) coming from $\beta\sp{2}\partial\sb{\beta}
F\sb{\beta}$, no matter how we choose fixed values of the masses
$m\sb{s}, m\sb{d}$ and couplings $\xi\sb{s}$ and thus it seems that,
within our regularization procedure, Frolov Fursaev Zelnikov's idea 
does not works.

With regard of this point, it is worth noticing that in the {\em
local} $\zeta$ function computation of the black-hole entropy there is
a very clear distinction between ultraviolet and horizon divergences:
the former are cured already in the local quantities by the $\zeta$
function procedure,  while the latter arise only in global quantities
obtained integrating over the manifold local quantities with
non-integrable divergences at the horizon. In order to control this
divergences the integrations are stopped at a distance $r=\epsilon$
form the horizon.

Instead, in the global heat-kernel plus proper-time regularization
employed in \cite{frofurzel97a,frofurzel97b} no such
 distinction is available (see also \cite{Solo95a}, where the
ultraviolet cut-off $L\sp{-1}$ and the minimal distance form the
horizon $\epsilon$ are identified, $\epsilon=L\sp{-1}$). Indeed, if
the {\em global} $\zeta$ function is computed taking the Mellin
transform of the {\em global} heat kernel, then the theory seems not
to have any horizon divergence, in contrast with the results of others
local approaches \cite{ZCV96}. Obviously the {\em global} $\zeta$
function obtained through this way disagrees with the {\em global}
$\zeta$ function obtained by integrating the {\em local} one. This
disagreement between global approach and local approach is peculiar of
Euclidean conical manifolds.

We stress further that  the direct global approach produces an
apparently unphysical temperature dependence for $\beta \neq
\beta\sb{H}$ \cite{ZCV96} and it is not clear how to relate the global
quantities to local quantities as the renormalized stress tensor in
such an approach.

We finally notice that the expression of $\langle\phi\sb{s}
\sp{2}\rangle\sb{\beta\sb{H}}$
employed in \cite{frofurzel97a,frofurzel97b} is simply the {\em flat
space} result, as can be easily checked, and so it shows the usual
ultraviolet divergences, but it has no horizon divergence. As a
consequence, it seems to us that the method employed in
\cite{frofurzel97a,frofurzel97b} to compute $S\sp{BH}$ allows the
cancellation of the ultraviolet divergences, as possible with 
less assumptions, within the Rindler approximation, using the local
$\zeta$ function, but not the cancellation of the horizon divergences,
which do not appear in \cite{frofurzel97a,frofurzel97b}.

\par \section*{Acknowledgments}

We would like to thank S. Zerbini for useful discussion. V. M. would
like to thank R. Balbinot, in particular for valuable discussions, and
G. Venturi for their very kind hospitality at the Dept. of Physics of
Bologna University. The part of this work performed by V. M. has been
supported by funds of Department  of Physics of Bologna University.

\end{document}